\documentstyle[12pt,epsf]{article}
\epsfverbosetrue
\def\figlabel#1{\xdef#1{\thefigure}}

\def\figalign#1#2#3{
\begin{figure}
\centerline{
\hbox to 2.5truein{\vtop{\hsize=2.5truein\epsfxsize=6cm
\centerline{\epsfbox{#1} }
\caption[]{#3}
\figlabel{#2}
}}
}
\end{figure}
}

\def\be{\begin{equation}}
\def\ee{\end{equation}}
\def\bea{\begin{eqnarray}}
\def\eea{\end{eqnarray}}
\def\G{\Gamma}
\def\L{\Lambda}
\def\a{\alpha}
\def\b{\beta}
\def\g{\gamma}
\def\d{\delta}
\def\pa{\partial}
\def\t{\widetilde}

\begin{document}
\begin{titlepage}
\begin{flushright}
{ ~}\vskip -1in
CERN-TH/97-37\\
US-FT-9/97, UB-ECM-PF 97/03\\
hep-th/9703072\\
March 1997\\
\end{flushright}
\vspace*{20pt}
\bigskip
\begin{center}
 {\Large SOFTLY BROKEN $N=2$ QCD \\
   \bigskip
\Large WITH MASSIVE QUARK HYPERMULTIPLETS, I}
\vskip 0.9truecm

{Luis \'Alvarez-Gaum\'e$^{a}$,
Marcos Mari\~no$^{b}$ and 
Frederic Zamora$^{a,c}$.}

\vspace{1pc}

{\em $^a$ Theory Division, CERN,\\
 1211 Geneva 23, Switzerland.\\
 \bigskip
  $^b$ Departamento de F\'\i sica de
Part\'\i culas,\\ Universidade de Santiago
de Compostela,\\ E-15706 Santiago de Compostela, Spain.\\
  \bigskip
$^c$ Departament d'Estructura i Constituents de la Materia,
\\ Facultat de F\'\i sica, Universitat de Barcelona,\\ 
Diagonal 647, E-08028 Barcelona, Spain.}\\

\vspace{5pc}

{\large \bf Abstract}
\end{center}

We present a general analysis of all possible soft breakings
of $N=2$ supersymmetric QCD preserving the analytic properties
of the Seiberg-Witten solutions for the $SU(2)$ group 
with $N_f=1,2,3$ hypermultiplets. 
We obtain all the couplings
of the spurion fields in terms of properties of the
Seiberg-Witten periods, which we express 
in terms
of elementary elliptic functions by uniformizing the
elliptic curves associated to each number of flavors.
We analyze in detail the monodromy properties of the
softly broken theory, and obtain them 
by a particular embedding into a pure gauge theory 
with higher rank group. This allows to write explicit
expressions of the effective potential which are close
to the exact answer for moderate values of the supersymmetry
breaking parameters. The vacuum structures and phases
of the broken theories will be analyzed in the forthcoming
second part of this paper.


\end{titlepage}

\def\theequation{\thesection.\arabic{equation}}

\section{Introduction}
\setcounter{equation}{0}

This is the first of two papers where we analyze the general
possible soft breakings of $N=2$-rigid supersymmetry compatible
with the special properties of the Seiberg-Witten solution
for the $N=2$ generalization of QCD with up to four massive
flavours of quark hypermultiplets \cite{swone, swtwo}. We will also
discuss the straightforward generalization to other groups,
in particular $SU(N_c)$ \cite{kl, klt, ds, groups}.
In \cite{swone, swtwo} Seiberg and Witten
use the holomorphy
properties of the prepotential describing the low energy
action of $N=2$ supersymmetry with vector multiplets,
some beautiful physical arguments,  and a large colection
of self-consistency conditions to obtain the exact 
expression for the effective action up to two derivatives
with and without multiplets. The solution is described by
associating to each theory a Riemann surface of genus one 
which is determined in the Coulomb branch by the physical
scale of the theory $\Lambda_{N_f}$ and the hypermultiplet 
masses. In terms of a unique meromorphic
differential on this Riemann surface they are able to 
write down the prepotential and all the couplings of the
theory, and to determine the exact mass spectrum of 
BPS states (as the original solution will be crucial
in what follows, we will present a brief summary of the
relevant results of \cite{swtwo} including
the analyticity and monodromy properties
of their solutions at the beginning of section
two).  Since so much information follows from the
solution of the pure $N=2$ one would like to be able
to break $N=2$ supersymmetry down to $N=0$ in terms
of a collection of soft breaking terms preserving
all the holomorphy and monodromy properties which
were crucial to determine the solutions in \cite{swone, swtwo}.
This was begun in \cite{soft, hsuangle} (for $N=1$ similar soft 
breaking analysis 
was done in \cite{softone, softwo} to understand the implications 
of the non-perturbative
superpotentials obtained by Seiberg and 
collaborators \cite{none, nadual} 
in the absence of supersymmetry).
In this paper we study the most general soft breakings
in $N=2$ QCD-like theories including massive
hypermultiplets.  We find that for a theory with
$N_f$ hypermultiplets there are $3(1+N_f)$
soft-breaking parameters satisfying all requirements
and they are associated with making the dilaton
a spurion (as in \cite{soft}) plus converting
the quark masses into spurion vector multiplets. 
This is analogous to gauging the quark number symmetries
and then freezing the vector multiplets to become
spurions.  Since each $N=2$-vector multiplet 
contains three real auxiliary fields, this gives
a $3(1+N_f)$-parameter space of softly broken
theories all preserving the holomorphy properties
of \cite{swone, swtwo}.  There are a number of interesting
features in the analysis of these theories and
their corresponding monodromies, and for this
reason we have preferred to collect all the
relevant information in this paper.  In the
second part \cite{AMZ2} we
will analyze the vacuum structure, patterns
of symmetry breaking, low energy Goldstone
boson Lagrangians and other interesting physical
issues which one would like to understand in
ordinary QCD but which here can be given 
explicit answers for moderate sypersymmetry
breaking parameters compared to the physical
scale of the theory.  We also postpone to
\cite{AMZ2} the analysis of possible
modifications of our results coming from
some recent proposals for the supersymmetric
effective action up to four derivatives
\cite{4deriv}. For the numerical
analysis of the effective potential for these
theories we need to have simple expressions
for the Seiberg-Witten solutions in the massive
cases in terms of controllable functions.  We 
obtain explicit expressions for these 
solutions in terms of elementary elliptic
integrals using the uniformization of 
elliptic curves by Weierstrass functions.

The structure of this paper is as follows:
In section two we begin by a quick summary
of some of the salient features of the
Seiberg-Witten analysis for gauge group
$SU(2)$ with up to three hypermultiplets.
In particular to each theory one can associate
a given elliptic curve and a unique abelian
differential of the third kind whose periods
parametrize the Seiberg-Witten solution
as a function of the modulus in the Coulomb
phase of the theory.  We then present a general
framework to obtain explicit expressions for
these periods by transforming the curves
to Weierstrass canonical form, and then
using uniformization.  This gives very explicit
results in terms of simple elliptic functions.
We test our expressions in several ways.
First the residues at their poles are given
by the expected linear combinations of 
the bare masses of the quark hypermultiplets,
second we verify that the weak coupling
behavior follows when the modulus goes to
infinity, and third, we verify that 
the BPS formula in terms of our expressions
leads to a vanishing monopole mass where
it is expected for $N_f=1,2,3$.  The explicit
expressions derived are crucial for the
numerical analysis in \cite{AMZ2}. In 
section three we analyze
the general form of soft breaking compatible
with monodromy invariance and holomorphy.  We 
present three different ways of looking at
the problem giving the same answer. We
can begin with the theory with masses, assume
that the masses are now given vector multiplets
associated with the gauging of quark numbers
plus the dilaton vector multiplets, and derive
the monodromy transformations associated to
the theory and all relevant dynamical variables.
We can also consider this theory as obtained 
from a Seiberg-Witten treatment of a 
$SU(2)\times U(1)^{N_f}$ theory in the infrared.
Since in this theory there are no ``baryonic"
monopoles, together with some simple physical
constraints we again obtain the properties derived
using the first method.  Finally, we consider
the theory with 
$SU(2)\times U(1)^{N_f}$ group as obtained 
by looking at the Seiberg-Witten curve for
$SU(2+N_p)$ \cite{kl}, 
with $N_p$ the number of poles of
the Seiberg-Witten abelian differential 
with $N_f$ massive hypermultiplets 
($N_p=1, 2, 4$, for $N_f= 1, 2, 3$, respectively), 
and study it in a particular
singular region of its moduli space. Namely,
to this group we associate a hyperelliptic
curve of genus $2N_p +1$, given by a polynomial
of degree $2N_p+3$. If its roots are
$e_1,e_2,e_3, r_1,s_1,r_2,s_2,...r_{N_p},s_{N_p}$
we can take the limit where $r_i,s_i$ 
coalesce precisely at the positions which 
are given by the poles of the Seiberg-Witten
differential in the massive case, and the
other three roots serve to determine the
embedding of $SU(2)$ in $SU(2+N_p)$.
This reproduces precisely the same results
as before including the monodromies.
We also present in section three an explicit
computation of the coupling matrices
of the spurion fields to themselves and
to physical fields because these are necessary
to determine the low-energy effective action.
We obtain these couplings by directly studying
the integrals involved. For $SU(N)$ the authors 
in \cite{dhoker}
have obtained similar expressions using the 
theory of integrable systems.  Here, since
we have a simple uniformization at hand, 
it is possible to derive the results directly.
Finally in section four we write down the
effective potential of the theory including 
all the spurion fields.  Here we present
the expressions that need to be analyzed
numerically to determine all the possible
vacuum structures of the different theories.
The result of this analysis will appear in
\cite{AMZ2}.

\section{$N=2$ QCD with Massive Quark Hypermultiplets}

\setcounter{equation}{0}

\subsection{The Seiberg-Witten Solution}

In this section we will focus on the features of the 
Seiberg-Witten solution which appear when bare masses for the 
hypermultiplets are included. For a general review of the 
work of Seiberg and Witten, see \cite{luis, lerche, bilal}. 

Matter hypermultiplets in the fundamental representation 
of $SU(N_c)$ can be given $N=2$ invariant bare masses through 
the $N=1$ superpotential 
\be
W_m=\sum_{f} m_f {\widetilde Q}_f Q_f,
\label{masas}
\ee
where $Q_f$, ${\widetilde Q}_f$ denote as usual the $N=1$ chiral multiplets 
corresponding to an $N=2$ hypermultiplet. In the case of a gauge group $SU(2)$, 
the flavour symmetry group $SO(2N_f)$ is explicitly broken by (\ref{masas}) 
down to an $U(1)^{N_f}$ subgroup, where the $U(1)$ for each hypermultiplet is 
the baryon number symmetry. For $SU(2)$ with $N_f\le 3$
 hypermultiplets in the fundamental representation, the 
low-energy effective action for $N=2$ QCD was determined in \cite{swtwo} 
for arbitrary masses. The explicit solution of the theory is encoded, as 
in the 
massless case, in a genus one algebraic curve, in such a way that $a$ and $a_D$ 
(the fields entering into the BPS mass formula) are obtained as contour 
integrals of an abelian differential defined on the curve. However, there are 
three remarkable differences with the massless case that give these theories 
their characteristic richness and complexity:

i) When bare masses for the quarks are introduced, the flavor symmetry group 
is generically broken to an abelian subgroup that can contribute to the 
central charge of the $N=2$ algebra. The central charge reads now
\be
Z=n_e a + n_m a_D + \sum_{f=1}^{N_f}S^f{m_f \over {\sqrt 2}},
\label{carga}
\ee
where $n_e$, $n_m$ and $S_f$ are respectively the electric, magnetic and 
baryonic quantum numbers. Notice that the physical electric charge $Q_e$ 
is different from the electric quantum number $n_e$ due to Witten's effect. 
In the same way, it has been recently shown that a similar phenomenon 
holds for the baryon number $S^f$ \cite{ferraprl}. In particular, the values 
of $S^f$ in the massive case can be different from the ones in the massless 
case. In \cite{ferraprl} this was argued from the point of view of the 
renormalization group flow from the massive to the massless theories, and in 
\cite{andreas} this was explicitly shown in the $N_f=1$ case. In principle, 
the solutions we give in the next subsection for $a$ and $a_D$ determine 
the values of the baryon quantum numbers in the massive case. These will be 
needed in the numerical analysis we will perform in a forthcoming paper.

ii) The Seiberg-Witten abelian differential $\lambda_{SW}$ is determined 
by the equation
\be
{\partial  \lambda_{SW} \over \partial u}={ {\sqrt{2}} \over 8 \pi} 
{dx \over y},
\label{swabel}
\ee
as this requirement guarantees the positivity of the metric. 
 The relation between $y$ and $x$ determines an elliptic     
curve (see below for explicit formulae).
In the massless case,  
$\lambda_{SW}$ is an abelian 
differential of the second kind.  
However, in the 
massive case, $\lambda_{SW}$ is of the third kind, ${ \it i.e.}$ it is a 
meromorphic one-form with single poles and non-zero residues. This implies 
that the quantities 
\be
a_D = \oint_{\alpha_1}\lambda_{SW}, \,\,\,\,\,\,\,\,\ 
a = \oint_{\alpha_2}\lambda_{SW},
\label{lineas}
\ee
where $\alpha_{1,2}$ are the homology one-cycles of 
the genus one algebraic curve, 
are not invariant under deformations of the cycles across the poles of 
$\lambda_{SW}$. The jumps in $a_D$, $a$ are of the form $2\pi i$ times the 
residue of $\lambda_{SW}$. In fact \cite{swtwo}, the residues are linear 
combinations of the bare masses. In the next subsection we will compute these 
residues for the $N_f \le 3$ and we will see that they have the expected 
structure. 

iii) As a consequence of the points above, the monodromy transformations 
of $a$, $a_D$ are no longer elements in $SL(2, {\bf Z})$. One must take into 
account the possibility of jumps in $a$, $a_D$ depending on the residues of 
$\lambda_{SW}$. The description of the duality structure requires now 
the introduction of an $N_f+2$-dimensional column vector $(m_f/{\sqrt{2}}, 
a_D, a)$ 
which transforms as 
\be
 \left(\begin{array}{c}
m_f/{\sqrt{2}}\\ a_{D} \\
                    a \end{array}\right) \rightarrow 
{\cal M}
\left(\begin{array}{c} m_f/{\sqrt{2}}\\a_D\\
               a \end{array}\right),
\label{monodro}
\ee
where 
\be
{\cal M}=\left(\begin{array}{ccc} 1 & 0 & 0 \\
                           p^f&\alpha& \beta \\
q^f & \gamma &\delta  \end{array} \right)
\label{matriz}
\ee
and 
$\alpha$, $\beta$, $\gamma$, $\delta$ form an $Sl(2, {\bf Z})$-matrix. 
The numbers $p^f$, $q^f$ are determined by the structure of the residues;
they are in general integer multiples of the baryonic charges. 
The invariance of the central charge (\ref{carga}) gives in addition the 
transformation properties of the quantum numbers appearing in (\ref{carga}). 
If we consider the $(N_f+2)$-dimensional row vector $W=(S^f, n_m, n_e)$, 
under a monodromy 
transformation it must change as $W \rightarrow W {\cal M}^{-1}$.   
  
\subsection{A General Procedure to Compute $a$, $a_D$}

In this subsection we will present a general 
procedure to compute $a$ and $a_D$ 
starting from the abelian differential $\lambda_{SW}$ of Seiberg 
and Witten, and we will present the results for the asymptotically free 
$SU(2)$ theories with massive matter hypermultiplets in the 
fundamental representation\footnote{When this paper was being written 
a paper appeared \cite{ferrari} 
where a similar approach is used in the $N_f=4$ theory,
and a paper in preparation by A. Bilal and 
F. Ferrari is announced where they analyze the $N_f \le 3$ 
theories with similar methods.}. For the theories 
with massless hypermultiplets, $a$ and $a_D$ can be computed 
using the Picard-Fuchs 
equations \cite{klt, ey, bf, isidro}, and the problem of 
integrating the Seiberg-Witten abelian differential along 
the two cycles of the torus is reduced to the problem of 
solving a differential
equation. However, when we introduce masses for the 
hypermutiplets, the 
Picard-Fuchs equations are of third order and the 
solutions are not easy to 
obtain with this procedure (although they have been obtained 
numerically in the semiclassical regime in \cite{ohta}). Explicit 
solutions have been obtained for the periods (the derivatives of 
$a_D$, $a$ with respect to  $u$) in \cite{masuda, andreastwo}. 
Our approach (which can be used 
in the massless
case as well) is to use Abel's theorem and the Jacobi 
inversion to uniformize 
the Seiberg-Witten curve. In this way the contour integrals defining 
$a$ and $a_D$ can be 
obtained in terms of elementary elliptic funcions.

As it is well known, every algebraic curve of genus one 
can be written in the 
Weierstrass form \cite{alg},
\be
Y^2=4X^3-g_2X-g_3=4(X-e_1)(X-e_2)(X-e_3), 
\label{weier}
\ee
where the coefficients $g_2$, $g_3$ are related to the 
roots $e_i$, $i=1, 2,3$ by the equations
\be
g_2=-4(e_2 e_3+e_3 e_1+e_1 e_2), \,\,\,\,\,\ g_3= 4e_1 e_2 e_3.
\label{ges}
\ee
To uniformize the curve, we use Abel's theorem, which 
states that an algebraic curve of genus one like (\ref{weier}) 
is of the form ${\bf C}/ \Lambda$ for some 
lattice $\Lambda \subset {\bf C}$. The map from 
${\bf C}/ \Lambda$ to the curve 
(\ref{weier}) is given by 
\be
\psi(z)=(\wp(z), \wp ' (z))= (X, Y), 
\label{abelmap}
\ee
where we consider $(X,Y)$ as inhomogeneous coordinates in ${\bf CP}^2$, 
and the 
Weierstrass function $\wp(z)$ verifies the differential equation
\be
(\wp ' (z))^2=4\wp(z)^3-g_2 \wp(z) -g_3.
\label{diff}
\ee
Under this correspondence, the half periods of 
the lattice $\Lambda$, $\omega_i /2$, $i=1,2,3$, 
$\omega_3 = \omega_1 + \omega_2$, 
are mapped to the roots $e_i=\wp (\omega_i/2)$ of the cubic 
equation in (\ref{weier}), and the 
differential $dz$ on ${\bf C}/\Lambda$ is mapped to 
the abelian differential 
of the first kind $dX/Y$. The map (\ref{abelmap}) has an inverse given by
\be
z=\psi^{-1}(p)=\int^{p}_{\infty}{ dX \over Y},
\label{inver}
\ee 
which is defined modulo $\Lambda$. 
We can obtain an explicit expression for the inverse map 
(\ref{inver}), doing 
the change of variable $t^2=(e_2-e_1)/(X-e_1)$, to obtain
\be
z= -{1 \over {\sqrt{e_2-e_1}}} F(\phi, k), 
\label{incom}
\ee
where $F(\phi, k)$ is the incomplete elliptic integral of the first kind, with 
modulus $k^2=(e_3-e_1)/(e_2-e_1)$, and 
${\rm sin}^2\phi = (e_2-e_1)/(\wp(z)-e_1)$.

In fact, all these functions can be computed in terms of the roots 
$e_i$ and elliptic functions. First of all 
we have the periods 
of the abelian differential $dX/Y$. We take the branch cut on the $X$-plane 
from $e_1$ to $e_3$, and from $e_2$ to infinity (see fig. 1), so that 
the $\alpha_1$ and $\alpha_2$ periods are given 
by
\bea
 \omega_1 &= & \oint _{\alpha_1}{ dX \over Y} = \int_{e_2}^{e_3} {dX \over 
\sqrt{(X-e_1)(X-e_2)(X-e_3)} },\nonumber \\ 
\omega_2 &= & \oint _{\alpha_2}{ dX \over Y}=\int_{e_1}^{e_3} {dX \over 
\sqrt{(X-e_1)(X-e_2)(X-e_3)}}.
\label{periodos}
\eea
Introducing now the complementary modulus $k'^2=1-k^2$, we 
obtain a representation of the periods in terms of the complete 
elliptic integral of the first kind,
\bea
\omega_1 &= & {2 i \over {\sqrt{e_2-e_1}}} K(k'),\\ \nonumber
\omega_2 &= & {2 \over {\sqrt{e_2-e_1}}} K(k).
\label{peris}
\eea

\figalign{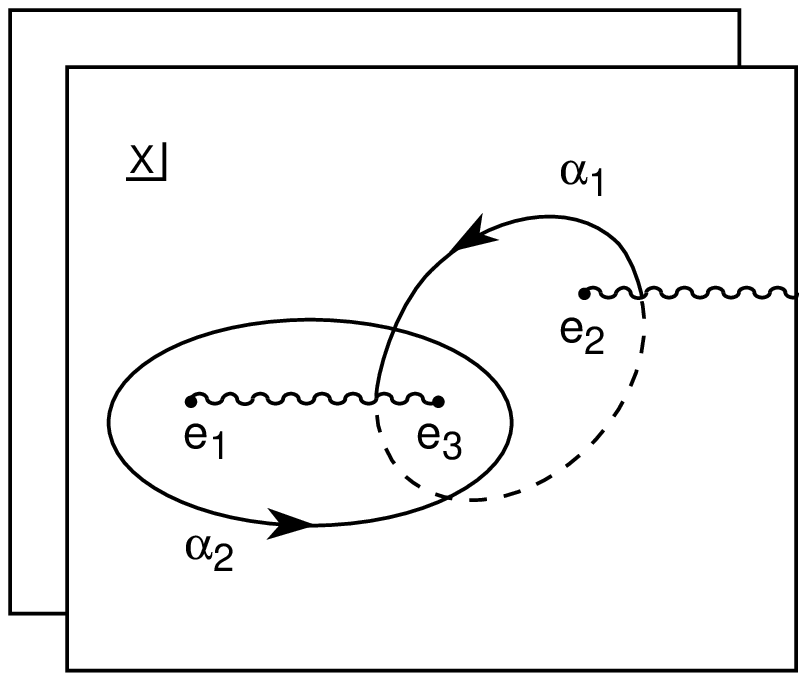}{\cyclesalpha}{The choice of the nontrivial one-cycles
$\a_1$ and $\a_2$.}

We will also need the Weierstrass $\zeta$-function, which is defined by 
the equation $\zeta ' (z)=-\wp (z)$. Because of this property, we have
that 
\be
\zeta ({\omega_i \over 2})=-\int_{\omega_j /2}^{\omega_3 /2} dz \wp (z),
 \,\,\,\,\,\,\,\,\,\ i, j=1,2, \ j\not=i,
\label{zetaper}
\ee
hence their values at the half periods 
can be computed in terms of complete elliptic
integrals,
\bea
\zeta ({\omega_1 \over 2}) &=& 
-i{ e_1 \over {\sqrt{e_2-e_1}}} K(k') - i 
{\sqrt{e_2-e_1}} E(k'), 
\\ \nonumber
\zeta ({\omega_2 \over 2}) &=&
 -{ e_2 \over {\sqrt{e_2-e_1}}} K(k) + {\sqrt{e_2-e_1}}  E(k).
\label{hypergen}
\eea
We can also give the value of $\zeta(z)$, for general $z$, in terms 
of incomplete elliptic integrals of first and second kind:
\be
\zeta(z)= \zeta({\omega_1 \over 2}) - {e_2 \over \sqrt{e_2-e_1}} 
F(\varphi, k) + \sqrt{e_2-e_1} E(\varphi, k),
\ee
where $\sin^2\varphi = (\wp(z) - e_1 )/(e_3-e_1)$.

As in the 
massive theories the Seiberg-Witten differential is of the third kind (with 
residues depending on the bare masses for the quark hypermultiplets), we need 
an expression for integrals on ${\bf C}/\Lambda$ of 
this kind of differentials. 
First of all we have the relation \cite{chandra, gra}
\be
-{\wp '(z_0)  \over {\wp (z) -\wp (z_0)}}= \zeta (z+z_0)+\zeta (z-z_0) -
2 \zeta (z_0).
\label{cociente}
\ee
We can now use the fact that the Weierstrass $\zeta$-function is the 
logarithmic derivative of the $\sigma$-function to obtain the expression
\be
\int {dz \over {\wp (z) -\wp (z_0)}} = {1 \over \wp ' (z_0)} 
\Bigl( 2z \zeta (z_0) 
+ {\rm ln} { \sigma (z-z_0) \over \sigma (z+z_0)}\Bigr). 
\label{indef}
\ee
As the $\sigma$-function is odd and verifies 
\be
\sigma ( z + \omega_i)=-\sigma(z) {\rm e}^{2 \zeta ({\omega_i \over 2}) (z 
+ {1 \over 2} \omega_i )}, 
\label{sigma}
\ee
we can obtain from (\ref{indef}) a simple expression 
for contour integrals of 
quotients like (\ref{cociente}) over the homology one-cycles:      
\be
\int_{\omega_i/2}^{\omega_3/2}{ dz \over {\wp (z) -\wp (z_0)}}= 
{1 \over \wp '(z_0)}\big( \omega_j \zeta(z_0) -
2 \zeta ({\omega_j \over 2}) z_0 \big).
\label{laforma}
\ee
Where $\omega_3=\omega_1+\omega_2$ and $i, j = 1, 2$, with $j\not= i$.
The Weierstrass function $\wp(z)$ is an even 
elliptic function of order
 two. This means that all the points $\pm z_0 + \Lambda$ 
give the same value to the Weierstrass function. 
Using Legendre's relation,
\be
\omega_1 \zeta({\omega_2 \over 2}) - \omega_2 \zeta({\omega_1 \over 2}) =
i\pi,
\label{Legendre}
\ee
one can verify that if we substitute $z_0$ by  
 $\pm z_0 + n_1\omega_1 + n_2\omega_2$, with $n_1, n_2$ integer numbers,
 in the 
integral (\ref{laforma}), we just obtain the additional term
$\pm (-1)^i 2 \pi i n_i / \wp'(z_0)$. On the other hand, 
notice that $-1/\wp '(z_0)$ 
is precisely the residue of $1/(\wp (z) -\wp (z_0))$.
As a contour integral, (\ref{laforma}) is then defined 
up to $\pi i n$ times the residue of $1/(\wp (z) -\wp (z_0))$, where $n$ 
is an integer. We will see that this is 
the behaviour we need in order to reproduce the jumps in $a_D$, $a$ that 
one expects in 
$N=2$ QCD with massive hypermutliplets. Also notice that the ambiguity 
in the choice of $z_0$ has the same structure that the one coming from the 
residue of the pole.    

When one of the periods of the lattice goes to infinity, say $\omega_2$ (this 
is precisely the case in the strong coupling singularities), one can derive 
expressions for the roots $e_i$ and $\zeta(\omega_1 / 2)$ in terms of 
$\omega_1$:
\be
e_3=e_2=-{e_1 \over 2}=-{\pi^2 \over 3 \omega_1^2 },\,\,\,\,\,\,\,\ 
\zeta({\omega_1 \over  2})={\pi^2 \over 6 \omega_1}.
\label{fuerte}
\ee
These are all the expressions we need to give explicit formulae for 
$a$ and $a_D$. 

\subsection{$a$, $a_D$ for $N=2$ $SU(2)$ with $N_f \le 3$}

We will now use the approach previously described to 
obtain explicit expressions 
for $a$ and $a_D$ for the $N=2$ $SU(2)$ gauge theory with 
$N_f \le 3$ hypermultiplets. 
To do this, we will use the cubic curves presented in \cite{swtwo}.
 As it should be clear from the discussion above, 
the first thing to do is to write 
the Seiberg-Witten curves in the Weierstrass form. 
Consider then the general 
form of the curves for the massive theories as described in \cite{swtwo}:
\be
y^2=x^3+ Bx^2+Cx +D,
\label{curvasw}
\ee
where the coefficients $B$, $C$ and $D$ depend on the gauge-invariant 
parameter 
$u$, the dynamical scale of the theory $\Lambda_{N_f}$, and on the bare 
quark masses. To put this curve in the Weierstrass form, 
it suffices to redefine the variables as 
\be
y=4Y, \,\,\,\,\,\,\,\ x=4X-{1 \over 3}B,
\label{cambio}
\ee
and the curve (\ref{curvasw}) has now the form given in (\ref{weier}) with
\be
g_2= -{1 \over 4}\big( C-{1\over 3}B^2),\,\,\,\,\,\,\,\ 
g_3=-{1 \over 16}\big(D+ {2 B^3 \over 27}-{CB\over 3}). 
\label{swges}
\ee
Notice that with the redefinition given in 
(\ref{cambio}), the abelian differential of the first 
kind is $dX/Y=dx/y$.

\subsubsection{$N_f\leq 3$ Massless}

 Before analyzing the three massive cases of the $SU(2)$ theory,
we give a review of the massless case within our approach. 
For the $SU(2)$ gauge theory with $N_f$ 
massless hypermultiplets, the Seiberg-Witten
abelian differential is \cite{iy}
\be
\label{SWabmassless}
\lambda_{SW} = {{\sqrt 2} \over 8\pi}{dx \over y} (2u - (4-N_f)x).
\ee
By the uniformization method, the integrals to do are 
($a_D$ and $a$ will be denoted by $a_1$, $a_2$, respectively):
\be
a_i = {\sqrt 2 \over \pi} \int_{\omega_j/2}^{\omega_3/2}
 dz \Big( ({N_f + 2 \over 12})u 
- {\d_{N_f,3} \Lambda^2_3 \over 768} - (4-N_f) \wp(z) \Big).
\ee
And just using formula (\ref{zetaper}) and that
$\zeta({\omega_3 \over 2}) = \zeta({\omega_1 \over 2}) + 
\zeta({\omega_2 \over 2})$, we obtain
\be
\label{amassless}
a_i=  {{\sqrt 2} \over \pi} \Big( (4-N_f)\zeta({\omega_i \over 2}) +
 ({ N_f+2 \over 24})u \, \omega_i - ({ \d_{N_f, 3} 
\Lambda_3^2 \over 1536}) \omega_i \Big)
\ee
In particular we immediately obtain the identity \cite{matone, 
sonn, ey}
\be
a \Big({da_D \over du} \Big) - a_D \Big({da \over du} \Big) = 
{i(4-N_f) \over 4\pi}
\ee
which holds for the massless case, just by direct application of the 
Legendre's relation (\ref{Legendre}).
From the analysis of the periods (\ref{periodos}) at weak coupling, 
\bea
\omega_1 &=& {i(4-N_f) \over {\sqrt u}}\Big( \log {u \over \Lambda^2_{N_f}}
 + 2 + O\Big({\Lambda^2_{N_f}\over u}\Big) \Big),
\\
\omega_2 &=& {2 \pi \over {\sqrt u}} \Big( 1 + 
O\Big({\Lambda^2_{N_f} \over u}\Big) \Big),
\eea
we have
\bea
\zeta({\omega_1 \over 2}) &\sim& {i(4-N_f) \over 24} {\sqrt u} 
\log {u \over \Lambda^2_{N_f}} - {i(2+N_f) \over 12} {\sqrt u};
\\
\zeta({\omega_2 \over 2}) &\sim& {\pi \over 12} {\sqrt u}.
\eea
Where we have used the homogeneity relation 
of the Weierstrass $\zeta$-function,
\be
\zeta(\lambda z; \lambda \omega_i) = \lambda^{-1} \zeta(z; \omega_i),
\ee
and the fact that $\zeta(\pi) = \pi / 12 $ for a lattice of periods 
$(2\pi, \infty)$,
to compute the asymptotic expression of $\zeta(\omega_2 / 2)$ in the
 weak coupling region.
By the Legendre's relation (\ref{Legendre}) we have obtained the 
behaviour of  $\zeta(\omega_1 / 2)$ in this region.
Substituting these expressions in 
(\ref{amassless}), we obtain  the expected weak coupling expressions
\bea
a &=& {\sqrt{2u} \over 2} \Big( 1 + O\Big({\Lambda^2_{N_f} \over u}\Big) \Big),
\\
a_D &=& {i(4-N_f) \over 4\pi} \sqrt{2u} \Big( \log {u \over \Lambda^2_{N_f}} +
 O\Big({\Lambda^2_{N_f} \over u}\Big) \Big).
\eea

Finally, we show a strong coupling test. 
On the strong coupling singularities,
$\omega_2$ diverges and we have, using (\ref{fuerte}), that
\be
a_D =  {{\sqrt 2} \over \pi} \Big( (4-N_f){\pi^2 \over 6 \omega_1} +
 ({ N_f+2 \over 24})u \, \omega_1 - ({\d_{N_f, 3} \over 64}) 
\omega_1 \Big),
\label{fuerteaD}
\ee
where we have gone to adimensional units, 
choosing $\Lambda^{(4-N_f)}_{N_f} = 8$. 
For instance, on the monopole singularity, we have that 
$$(u, \omega_1) = ( e^{-{i\pi \over 3}}, 2\pi e^{2\pi i \over 3}), \, (1, 
{2\pi i \over \sqrt 2}) , \, (0, -2\pi)$$
for $N_f=1, 2, 3$, respectively.
With these values (\ref{fuerteaD}) implies that 
\be
a_D = 0,
\ee
as expected.

\subsubsection{$N_f=1$ Massive}
The Seiberg-Witten curve is in this case
\be
y^2=x^2(x-u)+{1\over 4}m \Lambda_1^3 x -{1 \over 64}\Lambda_1^6,
\label{uncurva}
\ee
and with the redefinitions in (\ref{cambio}), $x=4X+ u/3$. This 
corresponds to a 
curve in the Weierstrass form (\ref{weier}) with 
\bea
g_2(u,m)&=&{1\over 4}\Big ( {u^2\over 3} - {1\over 4} m \Lambda_1^3 \Big),
\nonumber \\
g_3(u,m) &=& {1 \over 16}\Big( -{1\over 12}m u \Lambda_1^3+
{1 \over 64}\Lambda_1^6+{2\over 27}u^3 \Big).
\label{unges}
\eea
An explicit representative of the Seiberg-Witten differential 
can be easily obtained in this case:
\be
\lambda_{SW}= -{\sqrt{2} \over 8 \pi}{dx \over y}(3x-2u + 
{m\Lambda_1^4 \over 4x}),
\label{unab}
\ee
up to an exact differential. This differential has a pole at $x=0$, with 
residue
\be
{\rm Res}_{x=0} \lambda_{SW} = - {1 \over 2 \pi i}{m \over {\sqrt 2}},
\label{unres}
\ee 
in the positive Riemann sheet ({\it i.e.} for the positive 
sign of the square root).
To obtain $a_D$ and $a$, we will simply use the correspondence between 
$X$ and $\wp(z)$ given by the Abel map in (\ref{abelmap}), and the integrals 
(\ref{zetaper}) and (\ref{laforma}). Notice that the pole at $x=0$ 
corresponds to a pole at $z_0$, verifying 
$4\wp (z_0) + u/3=0$. In the case at hand, we can easily compute 
the residue of this pole, $1/ \wp ' (z_0)$, 
taking into account that $Y$ corresponds to $\wp ' (z)$ 
under the Abel map (\ref{abelmap}). In this way we obtain
\be
16 (\wp ' (z_0))^2= y^2(0)= -{1 \over 64} \Lambda_1^6, 
\label{unpes}
\ee
hence $\wp ' (z_0)= i \Lambda_1^3/32$, where the point $z_0$
 corresponds to the pole $x=0$ which lives in the positive Riemann sheet.
 The final expression for the 
$a_i$ is 
\be
a_i = { {\sqrt{2}}\over 4 \pi} \Bigg( 12 \zeta \big({\omega_i \over 2} \big) 
+u{\omega_i \over 2} + 2i m 
\big[\omega_i \zeta (z_0)-
2 z_0 \zeta \big({\omega_i \over 2} \big)  \big] \Bigg).
\label{unas}
\ee
Notice that the coefficient of the term 
in square brackets in (\ref{unas}) is precisely 
twice the residue of $\lambda_{SW}$ (with a positive sign in the 
square root). This is a general fact for all the expressions we will 
find below, and is 
a direct consequence of the structure of the abelian 
differential of third kind. We have seen that the integral 
(\ref{laforma}) is defined up to $\pi i n$ times $1/\wp ' (z_0)$. This 
implies that $a_D$, $a$ will be defined up to $2 \pi i n$ times 
the residue of $\lambda_{SW}$. This is the expected structure of the 
constant shifts in the monodromy transformations, as we recalled
 in the previous 
subsection following \cite{swtwo}. 
One should be careful with the global structure of the solutions 
on the $u$-plane. In general one must analytically 
continue the expressions of 
$a_D$, $a$ through the cuts. This simply amounts, in this formalism, to 
a permutation of the roots of (2.22) for each case
and they are given by the Weierstrass function at the half-periods
of the torus.  These issues will be discussed in the forthcoming
\cite{AMZ2} second part of this paper.

\subsubsection{$N_f=2$ Massive}

In this case the explicit expressions are slightly more involved. The 
Seiberg-Witten curve is given by
\be
y^2=(x^2-{1 \over 64} \Lambda_2^4)(x-u)+ {1 \over 4}m_1m_2 \Lambda_2^2x- 
{1 \over 64} (m_1^2 + m_2^2)\Lambda_2^4, 
\label{doscurva}
\ee
which can be written in the Weierstrass form with $x=4X+ {u \over 3}$ and 
coefficients
\bea
g_2(u,m_1,m_2) &=& {1 \over 4} \Big( {u^2 \over 3}+ {1 \over 64} \Lambda_2^4 -
{1 \over 4} m_1 m_2 \Lambda_2^2 \Big), 
\nonumber \\
g_3(u,m_1,m_2) &=& {1 \over 16} \Big( {2 u^3 \over 27} + {1 \over 64} 
(m_1^2 +m_2^2-u)\Lambda_2^4 \nonumber\\
&-& {u \Lambda_2^2 \over 12}
(m_1m_2 -{1 \over 16} \Lambda_2^2) \Big).
\label{dosges}
\eea
The abelian Seiberg-Witten differential can be easily computed in this case 
\cite{swtwo}:
\be
\lambda_{SW}= -{{\sqrt{2}} \over 4 \pi}{dx \over y} 
\Bigg( x-u + { 1 \over 16} \Lambda_2^2 
\Big( {m_{+}^2 \over {x+ { 1 \over 8} \Lambda_2^2}} - 
{m_{-}^2 \over {x- { 1 \over 8}\Lambda_2^2}}\Big) \Bigg),
\label{dosab}
\ee
where we have defined $m_{\pm}=m_1 \pm m_2$. We see that there are poles 
at $x_{\pm}= \mp \Lambda_2^2 / 8 $, with residues 
\be
{\rm Res}_{x=x_{\pm}}\lambda_{SW}=\mp { 1 \over 4 \pi i} 
{m_{\pm} \over {\sqrt 2}},
\label{dosres}
\ee
for the positive Riemann sheet.
 These poles are mapped to the points $z_{\pm}$ on
 ${\bf C}$ given by 
$4 \wp (z_{\pm})+ u/3 -x_{\pm}=0$. Taking into account the structure 
of the curve in the Weierstrass form one can compute 
\be
\wp ' (z_{\pm})= { i \over 32} \Lambda^2_2 m_{\pm}, 
\label{dosprima}
\ee
where, again, the $z_{\pm}$ are chosen in order to map on the
 positive Riemann sheet. 
One then obtains, 
\bea
a_i &=& {{\sqrt 2} \over 4 \pi}
\Big( 8 \zeta \big({\omega_i \over 2} \big) 
+ { 2 u \over 3} \omega_i \\ \nonumber & & +i \Big[ m_{+} \big( 
\omega_i \zeta (z_{+})-
2 z_{+} \zeta \big({\omega_i \over 2}\big) \big) -m_{-} \big( 
\omega_i \zeta (z_{-})-
2 z_{-} \zeta \big({\omega_i \over 2}\big) \big)  \Big] \Big),
\label{dosas}
\eea
where we can see the same pattern as in the $N_f=1$ case.

\subsubsection{$N_f=3$ Massive}

This is the more involved case. The Seiberg-Witten curve is given by
\be
y^2= x^2(x-u) - {1 \over 64} \Lambda_3^2 (x-u)^2 - {1 \over 64} 
\Lambda_3^2 t_2 (x-u) + {1 \over 4} \Lambda_3 t_3 x - {1 \over 64}
 \Lambda_3^2t_4, 
\label{trescurva}
\ee
where $t_2$, $t_3$, $t_4$ denote the polynomials
$$
t_2=m_1^2+m_2^2+ m_3^2, \,\,\,\,\,\ t_3= m_1m_2m_3, 
$$
\be
t_4= m_1^2m_2^2+m_2^2m_3^2+ m_1^2m_3^2.
\label{tes} 
\ee
The curve (\ref{trescurva}) can be written in the Weierstrass form with 
\be
x=4X+ {1 \over 3}\Big(u+ {1 \over 64} \Lambda_3^2 \Big).
\label{tresuni}
\ee

The computation of the appropriate Seiberg-Witten differential is   
somewhat involved, and we will
explain it in some detail. First of all we write the 
curve in terms of the $u$-variable (since, because of (\ref{swabel}), one 
must integrate with respect to it). Letting $c=\Lambda_3/8$, 
we have:
\be
y^2= -c^2u^2+p(x) u +q(x), 
\label{otraforma}
\ee
where
\be
p(x)= -x^2 + 2c^2 x+ t_2c^2,  \,\,\,\,\,\ 
q(x) = x^3 -c^2(x^2+t_2x+ t_4) + 2c t_3 x.
\label{polis}
\ee
Up to an exact differential, 
and a term which is $u$-independent (and then 
does not spoil the requirement in (\ref{swabel})), we obtain the 
explicit expression
\be
\lambda_{SW}= -{{\sqrt{2}} \over 16 \pi} {x dx \over y} 
{ {1 \over 2}pq' -p'q -u \Big( c^2q + {1 \over 4}p^2)' \over 
c^2q + {1 \over 4}p^2},
\label{tocho}
\ee
where the prime denotes derivative with respect to  $x$. 
The polynomial in the denominator has the roots
\bea
x_1 &=& { 1 \over 8} \Lambda_3(-m_1 + m_2 +m_3) ,
\\ \nonumber
x_2 &=& { 1 \over 8} \Lambda_3(m_1 - m_2 +m_3),  \\ \nonumber
x_3 &=&{ 1 \over 8} \Lambda_3(m_1 + m_2 -m_3 ), \\ \nonumber
x_4 &=& { 1 \over 8} \Lambda_3(-m_1 - m_2 -m_3), 
\label{raices}
\eea
which are in one-to-one correspondence with the weights of the 
negative-chirality spinor 
representation of $SO(6)$. The abelian differential simplifies 
and becomes:
\be
\lambda_{SW}= -{{\sqrt{2}} \over 16 \pi} {dx  \over y} \Big(
2x-4u - \sum_{n=1}^4 { y_n x_n \over x-x_n}\Big),
\label{tresab}
\ee
where the coefficients $y_n$ verify $y^2(x_n)=-c^2 y_n^2$ and are given explicitly 
by 
\bea
y_1 &=& u-m_1m_2-m_1m_3 +m_2m_3-x_1,
\\ \nonumber
y_2 &=&u-m_1m_2+m_1m_3-m_2m_3 -x_2,\\ \nonumber
y_3 &=&u+m_1m_2-m_1m_3-m_2m_3-x_3,\\ \nonumber
y_4 &=&u+ m_1m_2+m_1m_3+m_2m_3-x_4. 
\label{coes}
\eea
As a partial check of our result, notice that when all the three masses are 
zero,  we recover
precisely the abelian differential of the massless case \cite{iy}. With this 
explicit expression, we can easily compute the residues of the abelian 
differential at the poles $x_n$:
\be
{\rm Res}_{x=x_n}\lambda_{SW}= {1 \over 8 \pi i}{ {\hat x}_n 
\over { \sqrt{2}}},
\label{tresres}
\ee
where ${\hat x}_n= x_n / c$. The computation of $a$, $a_D$ is now straightforward. The poles $x_n$ 
correspond 
to the poles $z_n$ in ${\bf C}$ through the equation (\ref{tresuni}), 
and one easily obtains $\wp ' (z_n)=  i c y_n/4$. The final result is
\be
a_i = { {\sqrt{2}}\over 8 \pi} \Bigg( 8 \zeta \big({\omega_i \over 2} \big) 
+{\omega_i \over 3}\Big(5u- {1 \over 64} \Lambda_3^2 \Big) - i  
\sum_{n=1} ^4 {\hat x}_n \big[\omega_i \zeta (z_n)-
2 z_n \zeta \big({\omega_i \over 2} \big)  \big] \Bigg).
\label{tresas}
\ee

\section{Soft Breaking with Massive Quark Hypermultiplets}
\setcounter{equation}{0}

 In this section we study the general soft breaking of $N=2$ $SU(2)$ gauge 
theory with massive quark hypermultiplets, keeping the holomorphicity of the
low energy Lagrangian. 
In the first subsection we analyze the possibility 
 of promoting the hypermultiplets bare masses to the status of $N=2$ spurion
vector superfields. In the second subsection we give computable expressions
for the dual spurion masses and dilaton, and all the couplings.

\subsection{The Hypermultiplet Masses as $N=2$ Vector Superfields}

 The mass dependence of the Lagrangian (\ref{masas}) suggests 
the interpretation of the
 bare hypermultiplet masses as frozen $U(1)$ $N=2$ 
 vector multiplets of the baryonic flavour symmetries.
 Regarding them from this point of view,
 we could turn on their auxiliary fields obtaining additional supersymmetry
 breaking parameters. When the hypermultiplet masses are promoted
 to the status of
 abelian $N=2$ vector multiplets ${\cal M}_f$, we are really dealing with an
 effective Seiberg-Witten model with $SU(2) \times U(1)^{N_f}$ gauge group
 where the $N_f$ $U(1)$ factors come from the gauging of the
 baryon numbers\footnote{Effective model because the $SU(2) \times U(1)^{N_f}$
 gauge group does not have a good ultraviolet behaviour. At the end of
 the section 
we will interpret this gauge group as the low energy effective
gauge group of some asymptotically free gauge theory at some subdomain 
of its moduli space.}. At low energies, the $N=2$ effective Lagrangian of
 the vector multiplets includes the superfield ${\cal A}$ of the 
unbroken abelian subgroup of $SU(2)$, the $N_f$ mass spurions
 ${\cal M}_f$, with its scalar component giving the bare hypermultiplet masses
 $m_f$\footnote{In this section we rescale the quark masses by a
 $\sqrt{2}$ factor, and all repeated indices are summed.}, and 
the dilaton ${\cal S}$, whose scalar component
 gives the dynamically generated scale $\L= e^{is}$. The fact that
 the spurions are $N=2$ vector multiplets keeps the holomorphy
 of the leading part of the low energy Lagrangian, explicitly manifest
in $N=2$ superspace,
\be
\label{vmlagr}
{\cal L}_{\rm VM} = {1 \over 4\pi} {\rm Im}\Bigl( \int d^4 \theta
 {\cal F}({\cal A}, {\cal M}_f, {\cal S}) \Big).
\ee

 It is not possible to generate a ``magnetic'' baryonic charge, $S^f_D$,
 from the abelian baryonic factors, as the only 
states in the spectrum are ``electric''
baryonic ones, with baryonic charges $S_f$. 
Then, the central charge of the $N=2$ supersymmetry algebra of the
$SU(2) \times U(1)^{N_f}$ gauge group coincides with that of the $SU(2)$
gauge group with $N_f$ massive hypermultiplets (\ref{carga}). 
For a light $N=2$ BPS hypermultiplet $\{H, {\t H}\}$, with $(n_m, n_e, S^f)$
 charges, its effective Lagrangian is
\bea
\label{hmlagr}
{\cal L}_{\rm HM}&=& \int d^4 \theta \big( H^{*}{\rm e}^{2V' +2 S^f V_f}H +
 {\widetilde H}^{*}{\rm e}^{-2V'- 2 S^f V_f}{\widetilde H} \big)
\nonumber \\ 
 & + & \int d^2 \theta ({\sqrt 2} A' H {\widetilde H} +
{\sqrt 2} S^f M_f H {\widetilde H}) + {\rm h.c.},
\eea
where ${\cal A}' = n_m {\cal A}_D + n_e {\cal A}$.

The $N=1$ K\"ahler potential of (\ref{vmlagr}),
\be
\label{kahler}
 K(a, m_f, s;{\bar a}, {\bar m}_f, {\bar s}) =  {1 \over 4\pi} {\rm Im}
\Big( {\pa {\cal F} \over \pa a} {\bar a} + {\pa {\cal F} \over \pa m_f}
 {\bar m}_f + {\pa {\cal F} \over \pa s} {\bar s} \Big),
\ee
 because of its symplectic structure, is formally invariant under general
 $Sp(4 + 2 N_f, {\bf R})$ transformations $\G$,
\be
\left(\begin{array}{c} s_D \\ s \\ m^f_D \\ m_f \\ a_{D} \\
 a \end{array}\right) \rightarrow 
\left(\begin{array}{c} {\t s}_D \\ {\t s} \\ {\t m}^f_D \\
 {\t m_f} \\ {\t a}_D \\ {\t a} \end{array}\right) = 
\G  \left(\begin{array}{c} s_D \\ s \\ m^g_D \\ m_g \\ a_{D} \\
 a \end{array}\right);
\ee
where we have defined
\be
\label{spduals}
m^f_D = \left(\frac{\pa {\cal F}}{\pa m_f}\right)_{a, s}, \quad 
s_D =  \left(\frac{\pa {\cal F}}{\pa s}\right)_{a, m_f}.
\ee  
These transformations are the isometries of the
 corresponding K\"ahler metric, given by the imaginary part of
 the $(2 + N_f) \times (2 + N_f)$ matrix of couplings
\bea
\nonumber
\tau^{aa}= \frac{\partial^2 {\cal F}}{\partial a^2}, \quad
\tau^{fa}= \frac{\partial^2 {\cal F}}{\partial a \partial m_f}, \quad
\tau^{0a}= \frac{\partial^2 {\cal F}}{\partial a \partial s}, \quad
\\
\tau^{fg}= \frac{\partial^2 {\cal F}}{\pa m_f \pa m_g}, \quad
\tau^{0f}= \frac{\partial^2 {\cal F}}{\pa m_f \pa s}, \quad
\tau^{00}= \frac{\partial^2 {\cal F}}{\pa s^2}. 
\label{coupl}
\eea
 
The monodromy group of the $SU(2)\times U(1)^{N_f}$
model will be a subgroup of $Sp(4+2N_f, {\bf R})$. But there are 
some physical conditions that the monodromy
 transformations must fulfill, in order to preserve the 
structure of the BPS mass formula: 1) they should preserve the monodromy
invariance of the hypermultiplet masses and the dilaton,
 2) they should not generate a ``magnetic'' baryon number different from zero,
 3) they should not produce either ``electric'' or ``magnetic'' charges
for the dilaton vector multiplet, and 4) they should 
preserve the integrality of the electric and magnetic 
charges. If we
 restrict the action of $Sp(4 + 2N_f, {\bf R})$ to the subgroup that
 satisfies our physical inputs, we get that the possible monodromy
 transformations of the $SU(2)\times U(1)^{N_f}$ are:
\be
\label{monodr}
\left(\begin{array}{c} {\t s}_D \\ {\t s} \\ {\t m}^f_D \\
 {\t m_f} \\ {\t a}_D\\ {\t a} \end{array}\right) =
\left(\begin{array}{c} s_D \\ s \\ m^f_D + p^f(\g a_D + \d a)
 - q^f(\a a_D + \b a) + r^{fg} m_g \\ m_f \\ \a a_D+\b a +
 p^f m_f \\ \g a_D+\d a + q^f m_f \end{array}\right),
\ee
with $\left(\matrix{\a& \b \cr \g & \d \cr}\right) \in SL(2, {\bf Z})$,
 and $p^f, q^f, r^{fg} \in {\bf Q}$. The reason why 
$p^f,q^f,r^{fg}$ may
be rational instead of integers has to do with the possible
fractionalization of the quark numbers. For this reason
it is more convenient to state that the monodromy is
a subgroup of the group of symplectic rational matrices,
and in each case determine the subgroup by looking at
the explicit fractionalization (if any) of the quark numbers.
 
 Observe that $(a_D, a)$ transform as in the $SU(2)$ Seiberg-Witten 
theory with
 $N_f$ massive hypermultiplets (\ref{monodro}). 
This follows from conditions 2), 3) and 4) 
above. The fact that the transformation 
is an element of $Sp(4+2N_f,{\bf Q})$ 
fixes the monodromy transformations of $m^f_D$ and $s_D$. 
Notice that the dual dilaton spurion, $s_D$, is still a monodromy
 invariant,
as in the case of the theory with massless hypermultiplets \cite{soft}.
 But the dual mass spurions,
$m^f_D$, have a quite involved monodromy transformation. If the 
$SU(2)\times U(1)^{N_f}$ model with dilaton spurion is equivalent
to the $SU(2)$ theory with $N_f$ massive hypermultiplets, we must be able
to reproduce the same monodromy behaviour of $s_D$ and $m^f_D$ just from
the $SU(2)$ Seiberg-Witten solution with $N_f$ massive hypermultiplets.
 To check this, we should know the monodromy behaviour of the effective
 prepotential of massive $N=2$ QCD. It can be derived by integrating the 
identity 
\bea 
\Big(\frac{\pa {\cal {\t F}}({\t a}, m_f, s)}{\pa a} \Big)_{m_f, s}
 &=& \Big( {\pa {\t {\cal F}} \over \pa {\t a}} \Big)_{m_f, s}
 \Big(\frac{\pa {\t a}}{\pa a} \Big)_{m_f, s}
\nonumber
\\
&=& {\pa \over \pa a} \Big( {\cal F}(a, m_f, s) + {1 \over 2}
 \b \d a^2 + {1 \over 2} \a \g a^2_D + \b \g a a_D 
\nonumber
\\
 && + p^f m_f  (\g a + \d a_D) \big)|_{m_f, s},
\label{pretrans}
\eea
where we choose the $a$-independent constant to be zero.
On the other hand, from the Jacobian of the transformation $\{a, m_f, s\}
 \rightarrow \{\g a + \d + q^f m_f, m_f, s\}$ we obtain the relations
\bea
\Big( {\partial \over \partial {\t a}} \Big)_{\G -{\rm basis}}
&=&{1 \over \gamma \tau^{aa} +\delta }{\partial \over \partial a},
\label{deriva}
\\
\label{derivs}
\Big( {\partial \over \partial s} \Big)_{\G -{\rm basis}}
&=&{\partial \over \partial s}-{\gamma \tau^{a0} \over  \gamma
\tau^{aa} +\delta}{\partial \over \partial a},
\\
\Big( {\partial \over \partial m_f} \Big)_{\G -{\rm basis}}
&=&{\partial \over \partial m_f}-\frac{q^f+\g \tau^{af}}{\g \tau^{aa} +\d}
{\partial \over \partial a}.
\label{derivm}
\eea

Acting with  (\ref{derivs}) and (\ref{derivm}) on ${\t {\cal F}}$
 one obtains respectively the monodromy transformations of $m^f_D$ and $s_D$,
defined by equations (\ref{spduals}):
\bea
{\t m}^f_D &=& \Big( {\partial {\t {\cal F}} \over \partial m_f}
 \Big)_{\G -{\rm basis}} \nonumber 
\\
&=& m^f_D + p^f(\g a_D + \d a) - q^f(\a a_D + \b a) - q^f p^g m_g;
\label{mDt}\\
{\t s_D} &=& \Big( {\partial  {\t {\cal F}} \over \partial s}
 \Big)_{\G -{\rm basis}} = s_D.
\eea
Hence, we rederive the monodromy behaviour of $s_D$ in the  
$SU(2)$ massive theory\footnote{We will explicitly see that the
 expresion of $s_D$ turns out to be essentially 
the same as in the massles case.}. 
 The monodromy transformation for the dual mass of the 
$SU(2)\times U(1)^{N_f}$ model, 
given in (\ref{monodr}), coincides with that of the $SU(2)$ massive
theory in (\ref{mDt})
if we put $r^{fg}= -q^f p^g$. With this choice, the 
prepotentials of the $SU(2)\times U(1)^{N_f}$ model and the
$SU(2)$ massive theory are exactly
the same\footnote{Remember
 that there is an ambiguity in the definition of the 
 prepotential by terms quadratic in the vector multiplets,
 which allows to redefine $a_D$, $m^f_D$ and $s_D$ by linear terms
 in $a$, $m_f$ and $s$.}.

Finally, we can obtain the monodromy transformations of the couplings
(\ref{coupl})
acting with the derivatives (\ref{deriva}), (\ref{derivs}) and (\ref{derivm})
on $\{{\t a}_D, {\t m^f}_D, {\t s}_D \}$: 
\bea
{\t \tau}^{aa} &=& \frac{\a \tau^{aa} + \b}{\g \tau^{aa} + \d}, \quad
\nonumber\\
{\t \tau}^{0a} &=& \frac{\tau^{0a}}{\g \tau^{aa} + \d}, \quad
\nonumber\\
{\t \tau^{00}} &=& \tau^{00} - \frac{\g (\tau^{0a})^2}{\g \tau^{aa} + \d},
\nonumber\\
{\t \tau}^{af} &=& \frac{\tau^{af}}{\g \tau^{aa} + \d} -q^f (\frac{\a \tau^{aa} 
+ \b}
{\g \tau^{aa} + \d}) +p^f, \quad
\nonumber\\
{\t \tau^{f0}} &=& \tau^{f0} - (\frac{q^f+\g \tau^{fa}}{\g \tau^{aa} +\d})
 \tau^{0a},
\nonumber\\
{\t \tau^{fg}} &=& \tau^{fg} - \frac{\g \tau^{fa} \tau^{ga}}{\g \tau^{aa} + \d}
 - \frac{q^f \tau^{ga}}{\g \tau^{aa} + \d} - \frac{q^g \tau^{fa}}{\g \tau^{aa} + \d}
\nonumber\\
 & &- p^f q^g - p^g q^f + q^f q^g (\frac{\a \tau^{aa} 
+ \b}{\g \tau^{aa} + \d}).
\label{monotrans}
\eea

In this subsection we have shown the equivalence between the $N=2$ $SU(2)$ 
theory with $N_f$ massive hypermultiplets and a pure $N=2$ 
$SU(2) \times U(1)^{N_f}$ gauge model, supporting
 the interpretation of the bare quark masses as the scalar part
 of some frozen $N=2$ vector multipets ${\cal M}_f$. This opens the possibility
 of promoting them to the status of $N=2$ supersymmetry breaking spurion
 superfields, following the strategy of \cite{soft}, where the breaking was 
induced only by a dilaton spurion. At the same time we still keep the 
holomorphic structure of the low energy Lagrangian, as the $N=2$ prepotential 
has a holomorphic dependence on the hypermultiplet masses. 
Notice that, in contrast to $s_D$, the dual spurion mass
 is not a monodromy invariant. But this is not relevant as far as the 
$m^f_D$ dual spurion fields do not appear in
 the BPS mass formula. In other words, there are no ``magnetic'' baryonic 
charges associated to the monodromy invariant hypermultiplet masses.   
The important thing in this softly broken $N=2$ QCD
is that the extended monodromy transformations
(\ref{monodr}) belong to a subgroup of $Sp(4 +2 N_f, {\bf Q})$. This in 
turn guarantees the monodromy invariance of the vacuum energy 
induced by the supersymmetry breaking, since this energy is determined by the 
monodromy invariant $N=1$ K\"ahler potential (\ref{kahler}).

We have focused, just for simplicity, on the analysis
 of the soft breaking of $N=2$ $SU(2)$ with spurion masses and dilaton. 
 But the logic is straightforwardly
 extended for $SU(N_c)$ gauge groups with $N_f$ massive hypermultiplets
 ($N_f \leq 2 N_c$). 
The monodromy transformations of the $r=1, \cdots , N_c-1$ abelian vector 
multiplets of the $SU(N_c)$ gauge group are
\be
\Big(\begin{array}{c} a^r_D \\ a^r \end{array} \Big)
\rightarrow \Big(\begin{array}{c} {\t a}^r_D \\ {\t a}^r \end{array} \Big) =
\Big(\begin{array}{c} A^r_s a^s_D + B^r_s a^s + p^{rf} m_f \\
 C^r_s a^s_D + D^r_s a^s + q^{rf} m_f \end{array} \Big),
\label{genat}
\ee
where $\left(\matrix{A & B \cr C & D \cr}\right) \in Sp(2r, {\bf Z})$ and
$p^{rf}, q^{rf}$ are integer multiples of the baryonic charges. 
The effective prepotential has the monodromy behaviour
\bea
{\t {\cal F}}({\t a}^r, m_f) = &&{\cal F}(a^r, m_f) + {1 \over 2} 
a^T (D^TB)a + {1 \over 2} a^T_D (C^T A) a_D + a^T_D (C^T B) a 
\nonumber
\\
\label{genFt}
&&+ (p\cdot m)^T (C a_D + D a),
\eea
where $(p\cdot m)^r=p^{rf}m_f$. 
Computing the Jacobian of the monodromy transformations on the variables 
$\{a^r, m_f, s \}$, one obtains that
\bea
\Big( {\partial \over \partial {\t a}^r} \Big)_{\G -{\rm basis}}
&=&\Big[( C \tau + D )^{-1}\Big]^s_r {\partial \over \partial a^s},
\label{derat}
\\
\label{derst}
\Big( {\partial \over \partial s} \Big)_{\G -{\rm basis}}
&=&{\partial \over \partial s}- \Big[( C \tau + D )^{-1}\Big]^r_s
C^s_t \tau^{0t}
{\partial \over \partial a^r}
\\
\label{dermt}
\Big( {\partial \over \partial m_f} \Big)_{\G -{\rm basis}}
&=&{\partial \over \partial m_f}- \Big[( C \tau + D )^{-1}\Big]^r_s 
 (q^{sf} + C^s_t \tau^{tf} ) {\partial \over \partial a^r}.
\eea
where we define, as usual, the couplings
\bea
\nonumber
\tau^{rs}&=& \frac{\partial^2 {\cal F}}{\pa a^r \pa a^s}, \quad
\tau^{rf}= \frac{\partial^2 {\cal F}}{\partial a^r \partial m_f}, \quad
\tau^{r0}= \frac{\partial^2 {\cal F}}{\partial a^r \partial s}, \quad
\\
\tau^{fg}&=& \frac{\partial^2 {\cal F}}{\pa m_f \pa m_g}, \quad
\tau^{0f}= \frac{\partial^2 {\cal F}}{\pa m_f \pa s}, \quad
\tau^{00}= \frac{\partial^2 {\cal F}}{\pa s^2}. 
\eea
Acting with (\ref{derat}, \ref{derst}, \ref{dermt}) on (\ref{genFt}), 
one obtains the following monodromy transformations of the $N_f +1$ dual 
spurion fields:
\bea
{\t m}^f_D = \Big( {\partial {\t {\cal F}} \over \partial m_f}
 \Big)_{\G -{\rm basis}}  &=&  m^f_D + p^{rf}( C^r_s a^s_D + D^r_s a^s)
\nonumber
\\
 &-& q^{rf}( A^r_s a^s_D + B^r_s a^s) - q^{rf} p^{rg} m_g,
\label{genmDt}
\\
\label{gensDt}
{\t s_D} = \Big( {\partial  {\t {\cal F}} \over \partial s}
 \Big)_{\G -{\rm basis}} &=& s_D.
\eea
Notice that $s_D$ continues to be a monodromy invariant. 
One can easily verify that the monodromy transformations of 
the $(2N_c + 2N_f)$ dimensional vector 
$(s_D, s,  m^f_D, m_f, a^r_D, a^r)$ 
given by (\ref{genat}, \ref{genmDt}, \ref{gensDt}), with $s$ and $m_f$ 
invariants, are in a subgroup of the group 
$Sp(2N_c + 2N_f, {\bf Q})$ that leaves invariant the corresponding low energy 
K\"ahler potential. This guarantees the
monodromy invariance of the vacuum energy.

\subsection{Expressions for the Couplings}

In this subsection,
 we rederive some results of \cite{dhoker}, where 
an approach via integrable sistems was used.

To find the expression of $m^f_D$, we regard $a$ and $m_f$ as independent
variables and consider the second derivatives of the prepotential with 
respect to them:
\be
\Big({\pa m^f_D \over \pa a}\Big)_{m_g} = \Big({\pa a_D \over \pa m_f}\Big)_a
 = \oint_{\a_1} \Big({\pa \lambda_{SW} \over \pa m_f }\Big)_a.
\ee
The Riemann bilinear relation applied to the vanishing $(2, 0)$-form
$$\Big({\pa \lambda_{SW} \over \pa a }\Big)_{m_f} \wedge
\Big({\pa \lambda_{SW} \over \pa m_f }\Big)_a$$
gives
\be
\label{birel}
\oint_{\a_1} \Big({\pa \lambda_{SW} \over \pa m_f }\Big)_a =
2 \pi i \sum_{n=1}^{N_p} {\rm Res}_{x^{+}_n} [ \Big({\pa \lambda_{SW} \over
 \pa m_f }\Big)_a ]
\int_{x^-_n}^{x^+_n} \Big({\pa \lambda_{SW} \over \pa a }\Big)_{m_f}.
\ee

The points $x^+_n$ and $x^-_n$ ($n=1,\cdots,N_p$) are the simple
 poles of $\lambda_{SW}$
at each of the two Riemann sheets. Remember that the number of simple poles is
$N_p= 1, 2, 4$ for $N_f=1, 2, 3$, respectively. In (\ref{birel}) we have 
taken into account that
\bea
\oint_{\a_2} \Big({\pa \lambda_{SW} \over \pa m_f }\Big)_a &=& \Big({\pa \over
 \pa m_f }\Big)_a
\oint_{\a_2} \lambda_{SW} = 0;
\\
\oint_{\a_2} \Big( {\pa \lambda_{SW} \over \pa a }\Big)_{m_f} &=&
 \big({\pa \over \pa a }\Big)_{m_f}
\oint_{\a_2} \lambda_{SW} = 1.
\eea

\figalign{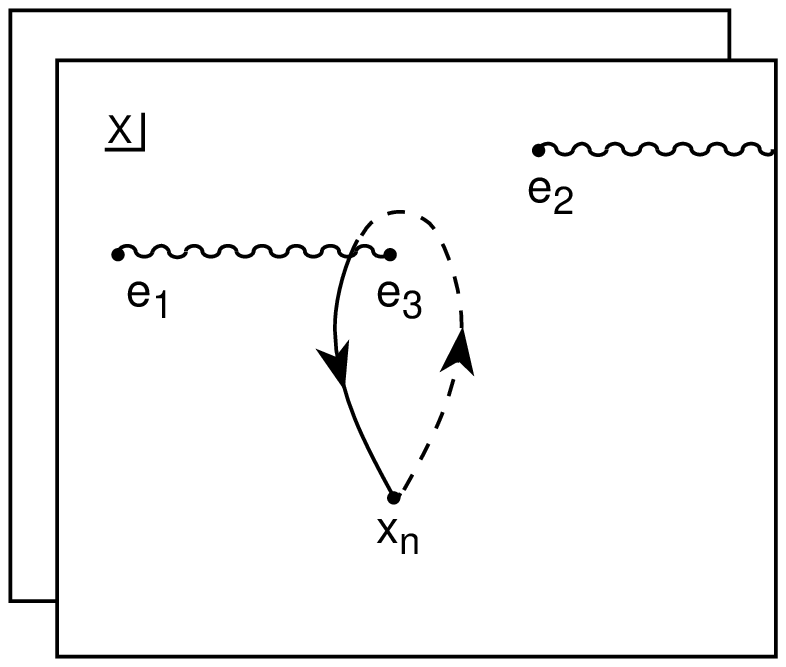}{\mDintegral}{The integral associated to 
the dual spurion masses.}

Since the poles $x_n$ and its corresponding residues 
are $a$-independent, the 
expression (\ref{birel}) can easily be integrated with respect to $a$, 
to obtain the formulae for the dual mass spurions (see fig. 2):
\be
\label{mD}
m^f_D = \sum_{n=1}^{N_p} S^f_n \int_{x^-_n}^{x^+_n} \lambda_{SW} +
 (a-{\rm{indep.\ const.}}),
\ee
where $S^f_n$ are $2\pi i$ times the residues in (\ref{birel}), 
{\it i.e.} the coefficients of the masses in the residues of 
$\lambda_{SW}$ corresponding to the poles $x_n$.
The resulting expression can be interpreted as an integral around
 some dual nontrivial one-cycle
associated to the pole $x_n$. To see this, consider the pure
 $N=2$ $SU(2 + N_p)$ gauge theory. The low energy description is 
encoded in a genus $1 + N_p$
 hyperelliptic curve described by a $(2N_p +3)$-order polynomial \cite{kl}. The 
roots of this polynomial will be denoted by ${\hat e}^n_1$, ${\hat e}^n_2$, 
$e_1$, $e_2$ and $e_3$, where $n=1, \cdots, N_p$. 
 We define 
\bea
{\hat m}_n = \oint_{{\hat \a}^n_2} {\hat \lambda}_{SW},
\\
{\hat m}^n_D = \oint_{{\hat \a}^n_1}{\hat \lambda}_{SW},
\eea
where ${\hat \lambda}_{SW}$ is the Seiberg-Witten abelian differential of the 
$SU(2 + N_p)$ theory, ${{\hat \a}^n_2}$ is a one-cycle going 
from ${\hat e}^n_1$ to ${\hat e}^n_2$, and ${{\hat \a}^n_1}$ is
the corresponding dual one-cycle, going from ${\hat e}^n_2$
to $e_3$. The remaining roots, $e_1$ and $e_2$, together with $e_3$, 
define the $a$ and $a_D$ variables of an embedded $SU(2) \subset SU(2+N_p)$ 
theory, with the conventions for the branch cuts as 
in section 2 (see fig. 3). 
Now imagine going to a singular 
submanifold of the moduli space of the $SU(2 + N_p)$ theory, where the roots 
${\hat e}^n_1$ and ${\hat e}^n_2$ coincide with the values $x_n$ of the poles 
of the massive $SU(2)$ theory, and 
${\hat \lambda}_{SW}$ goes to $\lambda_{SW}$. Then ${\hat m}_n$
and ${\hat m}^n_D$ will become $S^f_n m_f$ and $\int_{x^-_n}^{x^+_n}
 \lambda_{SW}$, respectively. In this region of the moduli space, we have 
$SU(2 + N_p) \rightarrow SU(2) \times U(1)^{N_f}$ and we recover 
the $SU(2)$ Seiberg-Witten model with $N_f$ massive
 hypermultiplets\footnote{Notice that the $S^f_n$ matrix has rank $N_f$.}.

\figalign{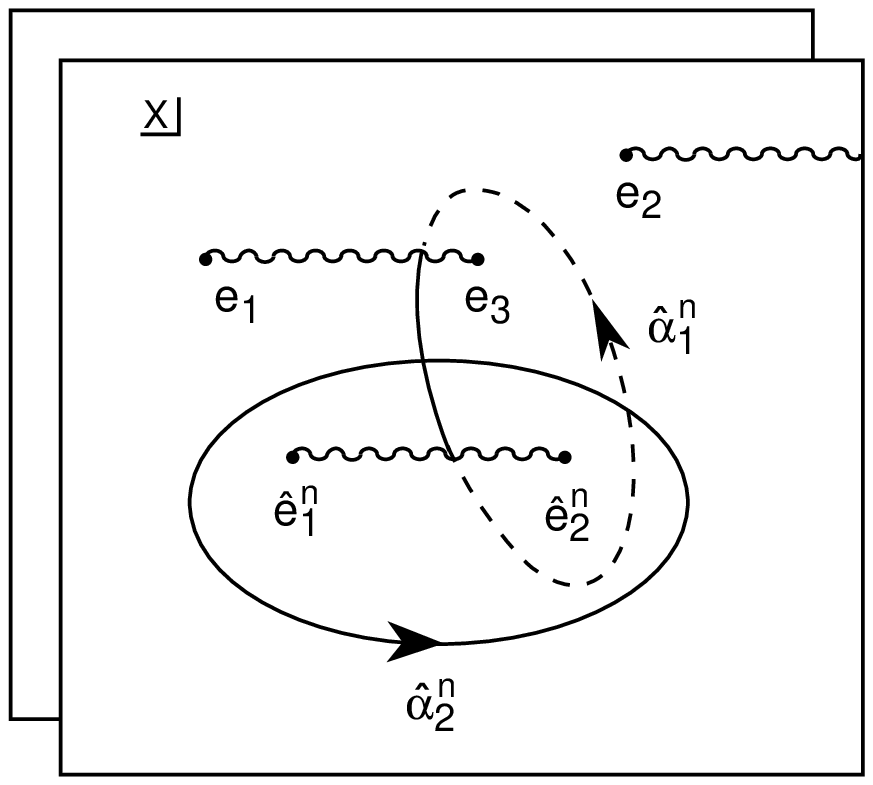}{\hatalpha}{The ${\hat \a}_1^n$ and ${\hat \a}_2^n$
one-cycles associated to the $x_n$ pole.}

The integral (\ref{mD}) can be computed by the uniformization method given
 in section 2,
but we must be careful with the integration limits. They are on the poles of 
$\lambda_{SW}$ and the integral diverges. This is why there are no 
``magnetic" baryonic charges appearing in the central charge of 
this $SU(2)$ massive theory embedded in a larger $SU(2+N_p)$ pure gauge theory.
 The corresponding ``baryonic monopoles"
 of the $SU(2 + N_p)$ theory have become infinitely massive in that 
singular region of the $SU(2 + N_p)$ moduli space, 
and they have been decoupled.
Effectively we should be able to substract this divergence
using the ambiguity in the prepotential with respect to the addition
of quadratic terms in $a$ and $m_f$.
This is possible because the residues of the
poles are only linear in the masses. 
We regularize the integral and focus 
on the part which gives the divergence:
\bea
{\rm div}(m^f_D) &=& \wp ' (z_n^+) {c_n \over 4} 
\int_{z^-_n - \epsilon}^{z^+_n + 
\epsilon} dz {1 \over \wp(z) - \wp(z^+_n)}
\nonumber 
\\
&=& c_n  
\Big( z^+_n \zeta(z^+_n) + z^+_n \zeta( {\omega_3
 \over 2}) - \Big({\omega_3 \over 2} \Big) \zeta(z^+_n)
\nonumber 
\\
 & & \,\,\,\,\,\,\,\,\,\ - \frac{1}{2} \log \sigma(2 z^+_n) + 
{1 \over 2} \log \sigma(\epsilon) \Big).
\eea
where we have chosen $z_n^+$ and $z_n^-=-z_n^+ + \omega_3$, both in the 
fundamental domain of the lattice. As we already know, the constant
$c_n$ appearing in the previous integral 
is just proportional to $S^f_n m_f$,
the corresponding residue of the $x_n$-pole. This means that we can
cancel the divergence just by a quadratic redefinition
of the prepotential. This is 
equivalent to choosing the $a$-independent 
constant in 
(\ref{mD}) to include $-{1 \over 2} c_n \log \sigma(\epsilon)$.  

Now we derive the expression of $s_D$ for massive $N=2$ QCD.
The Euler relation for the
prepotential gives 
\be
\label{homog}
2 {\cal F} = a a_D + m_f m^f_D + 
\Lambda {\pa {\cal F} \over \pa \Lambda}.
\ee
If we differentiate  with respect to the moduli parameter $u$, we obtain
\be
 {\pa \over \pa u}  \Big( m_f m^f_D + \Lambda {\pa {\cal F} \over
 \pa \Lambda} \Big) 
= \oint_{\a_2} {dx \over y} \oint_{\a_1} {\lambda}_{SW} -
 \oint_{\a_1} {dx \over y} \oint_{\a_2} {\lambda}_{SW}.
\ee
Applying  the Riemann bilinear relation to the vanishing $(2, 0)$-form 
$$
{dx \over y} 
\wedge {\lambda}_{SW},
$$ 
we have contributions from the residues of
the $x_n$ simple poles and from the constant factor 
$$
\pm {{\sqrt 2} \over 8\pi}
(4 - N_f)
$$ 
associated to the infinity points of the two Riemann sheets. The result is
\bea
 \oint_{\a_1} {dx \over y} \oint_{\a_2} {\lambda}_{SW} -
 \oint_{\a_2} {dx \over y} \oint_{\a_1} {\lambda}_{SW} 
\\
\,\,\,\,\,\,\,\,\,\,\,\ 
= { (4-N_f) \over 4\pi i } + \sum_{n=1}^{N_p} S^f_n m_f \int_{x^-_n}^{x^+_n} 
\Big( {\pa \lambda_{SW} \over \pa u} \Big).
\eea
Integrating with respect to $u$ and using (\ref{mD}), we identify 
\be
s_D= i \Lambda {\pa {\cal F} \over \pa \Lambda} = {(4-N_f) \over 4\pi} u 
\ + (a - \rm{indep. term}),
\label{sD}
\ee
showing its monodromy invariance.
The last expression is defined up to an integration constant. 
This integration constant will depend on the regularization of the 
effective prepotential to remove the divergence of (\ref{mD}).
This relation was obtained  in \cite{matone, sonn, ey} for the massless case; 
and in \cite{dhoker} for the massive case,
where concrete quadratic terms in the hypermultiplet masses appeared in the 
$a$-independent term of formula (\ref{sD}) because they chose a concrete
regularization of the prepotential. The choice of this integration
constant will be done by convenience in the numerical analysis of
the vacuum properties once supersymmetry is broken with soft terms
\cite{AMZ2}. 

Finally, we also give the explicit expressions for the couplings:
\bea
\tau^{fa} &=& \oint_{\a_1} \Big( {\pa \lambda_{SW} \over \pa m_f} \Big)_a
 = \Big( {\pa a_D \over \pa m_f} \Big)_u 
- \tau^{aa} \Big( {\pa a \over \pa m_f} \Big)_u,
\\
\tau^{0a} &=& \Big( {\pa s_D \over \pa a} \Big)_{m_f} = {(4-N_f) \over 4\pi }
 \Big( {\pa u \over \pa a} \Big)_{m_f},
\\
\tau^{0f} &=& {(4-N_f) \over 4\pi } \Big( {\pa u \over \pa m_f} \Big)_a
= - {(4-N_f) \over 4\pi } \Big( {\pa u \over \pa a} \Big)_{m_f}
\Big( {\pa a \over \pa m_f} \Big)_u,
\\
\tau^{00} &=& i ( 2 s_D - a \tau^{a0} - m_f \tau^{0f}),
\\
m_g \tau^{fg} &=& m^f_D - a \tau^{af} + i \tau^{0f},
\eea
where $\tau^{00}$ and $\tau^{fg}$ are obtained from the Euler relation
of the prepotential (\ref{homog}), acting with the appropriate derivatives.
Remember posible shifts linear in the masses for $\tau^{0f}$, coming 
from a concrete regularization of the prepotential.

\section{The Effective Potential}
\setcounter{equation}{0}

In this section we will present the computation of the effective potential 
when mass and dilaton spurions break both supersymmetries. 
We will denote the $N=1$ chiral multiplets corresponding to both the 
$U(1)$ gauge field and the spurions by $(A, S, M_f)$, 
$f=1, \cdots, N_f$,  and 
the corresponding $N=1$ vector multiplets by $(V_a, V_s, V_f)$. 
Roman capital letters 
refer to the mass and dilaton spurions only, 
$I,J=0,1,\cdots N_f$.  $N=2$ supersymmetry will 
be broken down to $N=0$ by turning on 
VEVs for the auxiliaries of the dilaton and mass spurions. In terms of 
$N=1$ superfields, we have:
$$
S=s + \theta^2 F^0, \,\,\,\,\,\,\,\ V_s= {1 \over 2} D^0 \theta^2 
{\bar \theta}^2, 
$$
\be
M_f= {m_f \over {\sqrt 2}}+ \theta^2 F^f,
 \,\,\,\,\,\,\,\ V_f= {1 \over 2} D^f \theta^2 {\bar \theta}^2,
\label{vevs}
\ee
where $m_f$ are the bare quark masses, and $s$ is related to the dynamical 
scale of the theory in the effective Lagrangian or to the classical 
gauge coupling in the bare Lagrangian. At the classical level, 
the breaking has 
the following effects. The dilaton spurion gives mass to the gauginos of 
the $N=2$ vector mutiplet and to the imaginary part of the 
Higgs field $\phi$,
the scalar component of the non-abelian $N=2$ vector multiplet 
\cite{hsuangle}; it also gives couplings between
the squarks and the Higgs field of the form
\be
{\overline F}^0 {\tilde q}_f ({\rm Im} \phi)q_f + {\rm h.c.}, 
\,\,\,\,\,\,\,\ D^0\Bigl(q^{\dagger}_f({\rm Im} \phi)q_f-
{\tilde q}_f({\rm Im} \phi) 
 {\tilde q}^{\dagger}_f \Bigr),
\label{squarks}
\ee
while the mass spurions give terms for the squarks with the structure
\be
F^f{\tilde q}_f q_f + {\rm h.c.}, \,\,\,\,\,\,\,\ 
D_f (|q_f|^2-|{\tilde q}_f|^2).
\label{mquarks}
\ee
In both cases, no new terms appear for the fermionic quarks. We see that the 
disadvantage of working with an $N=2$ spurion is 
that we do not generate a diagonal mass term for the squarks as in the soft 
breaking of $N=1$ supersymmetry. However, we have a 
better analytic control of the 
theory (at least for small supersymmetry breaking parameters). 
We will denote the couplings in the effective Lagrangian by 
\be
b_{IJ}= {1 \over 4\pi} {\rm Im} \tau_{IJ},
\label{couplings}
\ee
which are (up to a normalization) the components of the K\"ahler metric.
 As in section 2, we suppose that we are in the variables (electric, 
magnetic or dyonic) adequate to light BPS states. As all the 
states in the Seiberg-Witten singularities have the same charge (and 
normalized to 
one) with respect to  
the corresponding ``photon", we have to introduce only the baryon numbers 
appearing in the central charge (\ref{carga}). 
For $k$ light BPS states with $S^f_i$ baryonic charges, 
$i=1,\cdots, k$, the terms in (\ref{vmlagr}) and (\ref{hmlagr})
contributing to the effective potential are
\bea
V &=& b_{AB}F^A {\overline F}^B + 
b_{aA}\big( F^A{\overline F}^a+{\overline F}^A F^a \big) 
+b_{aa}|F_a|^2  \nonumber \\
&+& {1 \over 2} b_{AB} D^A D^B +  
+ b_{aA}D^aD^A+ {1 \over 2}b_{aa} D_a^2  \nonumber \\
&+& ( D_a +D_f S^f_i) (|h_i|^2-|{\widetilde h}_i|^2)+
 \sum_{i=1}^{k} \Big( |F_{h_i}|^2+ |F_{{\widetilde h}_i}|^2 \Big) 
\nonumber \\
&+& \sqrt{2} \Bigl( F^a h_i {\widetilde h}_i+
 a h_i F_{{\widetilde h}_i}+ 
a {\widetilde h}_i F_{h_i} + {\rm h.c.} \Bigr) \nonumber \\
&+& \sqrt{2} \Bigl( F^f S^f_i h_i {\widetilde h}_i+
 {m_f \over {\sqrt{2}}} S^f_i h_i F_{{\widetilde h}_i}+ 
{m_f \over {\sqrt{2}}} S^f_i {\widetilde h}_i F_{h_i} + {\rm h.c.}\Bigr),
\label{twov}
\eea
where all repeated indices are summed. We eliminate the auxiliary fields 
and obtain:
$$
D_a=-{1 \over b_{aa}}\Bigl( b_{aA}D^A + 
\sum_{i=1}^{k} (|h_i|^2-|{\widetilde h}_i|^2) 
\Bigr),
$$
$$
F_a=-{1 \over b_{aa}} \Bigl( b_{aA}F^A+
\sqrt{2}{\overline h}_i{\overline {\widetilde h}}_i\Bigr),
$$
\be
F_{h_i}=
-\sqrt{2}({\overline a}+ S^f_i {{\overline m}_f \over {\sqrt{2}}})
{\overline {\widetilde h}}_i,
\,\,\,\,\,\,\,\ 
F_{{\widetilde h}_i}=-\sqrt{2}({\overline a}+ S^f_i 
{{\overline m}_f \over {\sqrt{2}}})
{\overline h}_i.
\label{aux}
\ee 
Which substituted in (\ref{twov}) yields:
\bea
V&=&\Bigl({b_{aA} b_{aB} \over b_{aa}}-b_{AB}\Bigr) 
\Bigl({1 \over 2}D^A D^B + F^A {\overline F}^B \Bigr) + 
{b_{aA} \over b_{aa}}D^A \sum_{i=1}^{k} (|h_i|^2-|{\widetilde h}_i|^2) 
\nonumber \\ 
&+& { \sqrt{2} b_{aA}\over b_{aa}} \Bigl( F^A
  h_i{\widetilde h}_i + {\overline F}^A 
{\overline h}_i{\overline {\widetilde h}}_i \Bigr) + 
{2 \over b_{aa}} {\overline h}_i{\overline {\widetilde h}}_i
h_j{\widetilde h}_j \nonumber\\
&+& {1 \over 2 b_{aa}}\sum_{i,j=1}^{k} (|h_i|^2-|{\widetilde h}_i|^2)
(|h_j|^2-|{\widetilde h}_j|^2) -  
 D_f S^f_i (|h_i|^2-|{\widetilde h}_i|^2) \nonumber\\
&+& 2|a + S^f_i {m_f \over {\sqrt 2}}|^2(|h_i|^2+|{\widetilde h}_i|^2) 
\nonumber 
\\
&-&
\sqrt{2} \Bigl( S^f_i F^f h_i{\widetilde h}_i+ 
S^f_i{\overline F}^f {\overline h}_i{\overline {\widetilde h}}_i  \Bigr),
\label{twovi}
\eea
where $(b_{aA} b_{aB} / b_{aa})-b_{AB}$ is 
the cosmological term. This term in the 
potential must be a monodromy invariant, as we expect from the 
analysis in section 3. One can indeed check it explicitly using 
the monodromy transformations of the couplings given in (\ref{monotrans}).  
     
The potential (4.7) has a very rich structure of vacua
for the different flavors, and there are a number of
interesting questions in ordinary QCD that can be translated
to this context.  We will present the detailed analysis
in \cite{AMZ2}.

\section{Conclusions}

In this paper we have shown that it is possible to
softly break $N=2$-supersymmetric extensions of 
QCD in a way that preserves the analytic properties
of the solutions in \cite{swone, swtwo}.  We obtain a     
$3(N_f +1)$ space of parameters associated to the
auxiliary fields of $N_f+1$ spurion vector superfields.
One of them is the dilaton multiplet of $N=2$
supergravity, and the others are associated to
the gauging of the quark number symmetries for 
each hypermultiplet.  Although the analysis has
been carried out for the gauge group $SU(2)$, the
generalization is straightforward for other groups.
We have obtained explicit expressions for the
Seiberg-Witten periods using the standard uniformization
of elliptic curves, and verified our expressions
at strong, and weak coupling and by 
computing the residues of their poles which are linear
combinations of the bare quark masses.
An important ingredient in the determination of the
effective action in the presence of spurions is
the question of monodromy invariance.  We have
analyzed this issue by three different procedures
giving the same answer.  In particular we have
obtained our results by embedding the $SU(2)$ 
theory in the $SU(2+N_p)$ theory looking at
a singular subset of its moduli space where
some of its monopoles and dyon excitations have
infinite mass.  We have obtained explicit expressions
for the couplings of spurions to other multiplets,
and as a consequence we can write down the exact
form of the effective potential for moderate 
values of the supersymmetry breaking parameters
with respect to the dynamical scale of the theory.
It remains now to look at the possible vacuum
structures, patterns of symmetry breaking and
low energy phenomena encoded in this effective
action. This is currently under investigation.

It is clear that although the theories we have
treated are still far from real QCD, one can
nevertheless pose many analogous questions 
in these softly broken theories and obtain
explicit answers.  The issue is how one could
eventually compare with the real theory.
There are a number of obstacles that remain
to be dealt with, most notably the fact that
we  have a scalar field in the adjoint 
representation. Only in the limit when this
field decouples it seems possible to obtain
the structure of global symmetries in QCD. It
is difficult to see how to do this in the case
at hand, but it might be possible in some
large $N$ limit version of our results.
This as well as the possible effects
of higher derivative terms in the effective
action, and how they could modify the vacuum
structures, are being investigated.
We plan to report on our results elsewhere.

\newpage

{\large\bf Acknowledgements}

We acknowledge J.L.F. Barb\'on for valuable 
discussions and a critical
reading of the manuscript.
We also thank Isabelle Canon for helping with
the figures.  L.A.G. would like to thank the Physics
Department of Tokyo University where part of this work
was done. 
M.M. and F.Z. would like to thank the Theory 
Division at CERN
for its hospitality. 
The work of M.M.~is supported in part by DGICYT under grant
PB93-0344 and by
CICYT under grant AEN94-0928. The work of F.Z. is supported 
by a fellowship from Ministerio de Educaci\'on.

\end{document}